\documentclass[aps,prd,preprint,groupedaddress]{revtex4}

\usepackage{graphicx}
\usepackage{amsmath}
\usepackage{bm}

\def\ee{\end{eqnarray}}

\def\=:{=\hspace{-.7em}\raisebox{1.1ex}{.}\hspace{.1em}\raisebox{-0.2ex}{.} }

\newcommand {\beq}{\begin{eqnarray}}
\newcommand {\eeq}{\end{eqnarray}}

\newcommand {\1}[1]{\frac{1}{#1}}

\newcommand {\ph}{\varphi}

\newcommand {\del}{\partial}

\newcommand{\hs}[1]{\hspace{#1 mm}}

\begin{document}


\title{Knots from wall--anti-wall 
annihilations with stretched strings
}


\author{Muneto Nitta$^{1}$}

\affiliation{
$^1$Department of Physics, and Research and Education Center for Natural 
Sciences, Keio University, Hiyoshi 4-1-1, Yokohama, Kanagawa 223-8521, Japan\\
}


\date{\today}
\begin{abstract}
A pair of a domain wall and an anti-domain wall 
is unstable to decay.
We show that when a vortex-string is stretched between 
the walls, 
there remains a knot soliton (Hopfion) after the pair annihilation.

\end{abstract}

\maketitle

\section{Introduction}
The Faddeev-Skyrme (FS) model is 
an $O(3)$ model with a fourth derivative term.
As a skyrmion is a stable soliton 
in the Skyrme model \cite{Skyrme:1961vq}, 
the FS model admits a knot soliton, which is a stable soliton 
with a non-trivial Hopf number 
$\pi_3(S^2) \simeq {\bf Z}$ \cite{Faddeev:1996zj}. 
Such a soliton with the minimum Hopf charge 
is also called a Hopfion. 
A knot soliton, which is mathematically interesting 
in itself, is also considered as a model of a glue ball.  
Although several studies have been conducted on 
knot solitons, 
they have, thus far, focused most on 
soliton stability, construction of solutions, 
and interaction between multiple solutions,
among other topics.
However, thus far, there has been no discussion 
on how knot solitons are created. 
In contrast, a mechanism has already been proposed 
for the generation of topological defects:
the Kibble-Zurek mechanism, by which  
topological defects are inevitably 
generated during a phase transition 
accompanied with spontaneous symmetry breaking.

In this paper, we propose a mechanism 
of the creation of knot solitons. 
The key concept is the identification of a Hopfion 
as a closed and twisted baby skyrme string, {\it i.e.} 
a ``twisted baby skyrme ring".
A baby skyrmion \cite{Piette:1994ug} 
is a lump soliton in the ${\bf C}P^1$ model 
with a four-derivative term in 2+1 dimensions.
In 3+1 dimensions, it is a string-like soliton.
When one makes a closed baby skyrme string, 
one can twist the phase modulus.
Such a twisted baby skyrme ring is 
nothing but a Hopfion \cite{deVega:1977rk,
Kundu:1982bc,Gladikowski:1996mb}.
The original baby skyrme model 
admits the unique vacuum \cite{Piette:1994ug}, 
whereas the new baby skyrme model, 
proposed in a subsequent study \cite{Weidig:1998ii},
admits two discrete vacua.  
Therefore, the new model admits a domain wall solution 
interpolating between these two vacua
\cite{Kudryavtsev:1997nw,Harland:2007pb}. 
We consider a pair of a domain wall and 
an anti-domain wall. 
Such a configuration is clearly unstable 
to decay. 
However, it does not have to end up 
uniformly with the vacuum state. 
We first show that 
there can remain (anti-)baby skyrmions 
in $d=2+1$ dimensions 
and baby skyrme rings in $d=3+1$ dimensions. 
Untwisted baby skyrme rings are unstable to decay.  
However, when vortex-strings 
are stretched between the wall--anti-wall pair, 
there can appear twisted baby skyrme rings, 
or Hopfions.


\section{Faddeev-Skyrme Model}
The Lagrangian of the FS model is expressed as 
\cite{Faddeev:1996zj}
\beq
&& {\cal L} = \1{2} \del_{\mu}{\bf n}\cdot \del^{\mu} {\bf n} 
 - {\cal L}_4({\bf n})
 - V({\bf n}), \quad  
{\bf n} \cdot {\bf n} = 1 \label{eq:Lagrangian}
\eeq
with ${\bf n} = (n_1,n_2,n_3)$;
the fourth derivative FS term is expressed as
\beq
{\cal L}_4 ({\bf n})
= \kappa  \left[{\bf n} \cdot 
 (\partial_{\mu} {\bf n} \times \partial_{\nu} {\bf n} )\right]^2
= \kappa (\partial_{\mu} {\bf n} \times \partial_{\nu} {\bf n} )^2 , 
\eeq
in which we have used ${\bf n}\cdot \del {\bf n}=0$.
The original FS model has no potential, 
{\it i.e.}, $V=0$ \cite{Faddeev:1996zj}.
The original baby skyrme model has been proposed 
to have the potential $V= m^2 (1-n_3)$,  
which admits the unique vacuum $n_3=1$
\cite{Piette:1994ug}.
On the other hand, 
the new (or planar) baby skyrme model 
has been proposed with the potential 
$V= m^2 (1-n_3^2) = m^2 (1-n_3)(1+n_3)$ 
with two vacua $n_3 =\pm 1$ \cite{Weidig:1998ii}.
The latter admits both 
a baby skyrmion \cite{Weidig:1998ii} and a domain wall 
interpolating between the two vacua 
\cite{Kudryavtsev:1997nw,Harland:2007pb}. 

It is known that the $O(3)$ sigma model  
is equivalent to the ${\bf C}P^1$ model. 
To demonslate this, we use the relations
${\bf n} = \Phi^\dagger {\bf \sigma} \Phi$ and 
$\Phi^T = (1,u)/\sqrt{1 + |u|^2}$, 
where $u$ is the projective coordinate of ${\bf C}P^1$. 
The Lagrangian (\ref{eq:Lagrangian}) 
with the new baby skyrme potential 
can then be rewritten as 
\beq
{\cal L} =
 {\partial_{\mu} u^* \partial^{\mu} u - m^2 |u|^2 
  \over (1 + |u|^2)^2}  - 
2 \kappa {(\del_{\mu}u^* \del^{\mu}u)^2 
 - |\del_{\mu} u \del^{\mu} u|^2 
 \over (1+|u|^2)^4}. 
\eeq
Here, the two discrete vacua $n_3=+1$ and $n_3=-1$ are mapped 
to $u=\infty$ and $u=0$, respectively.

\begin{figure}[h]
\begin{center}

\includegraphics[width=0.2\linewidth,keepaspectratio]{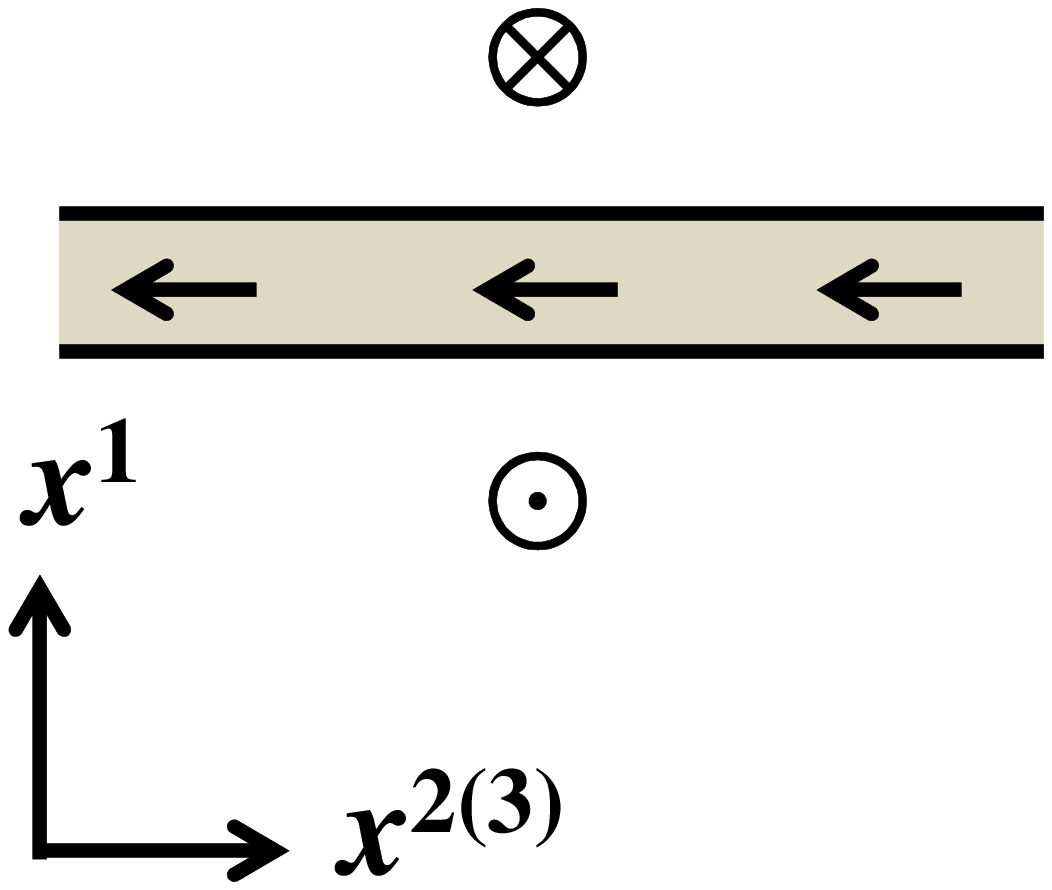}
\quad
\includegraphics[width=0.2\linewidth,keepaspectratio]{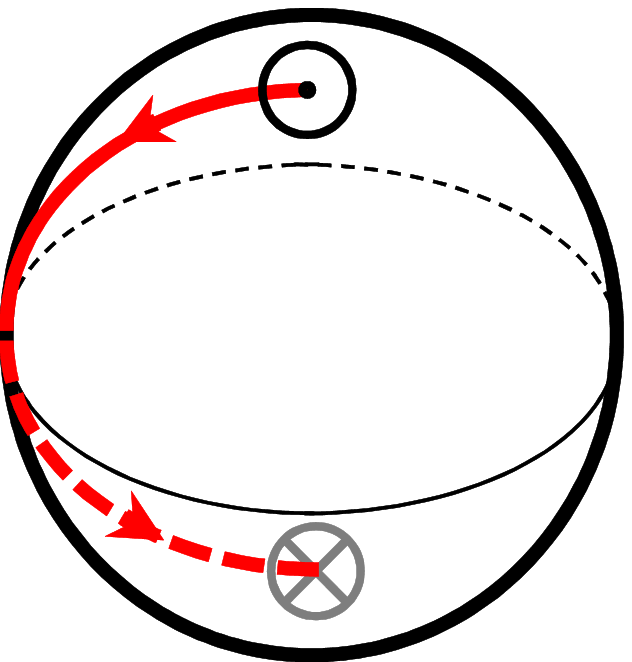}
\quad
\includegraphics[width=0.2\linewidth,keepaspectratio]{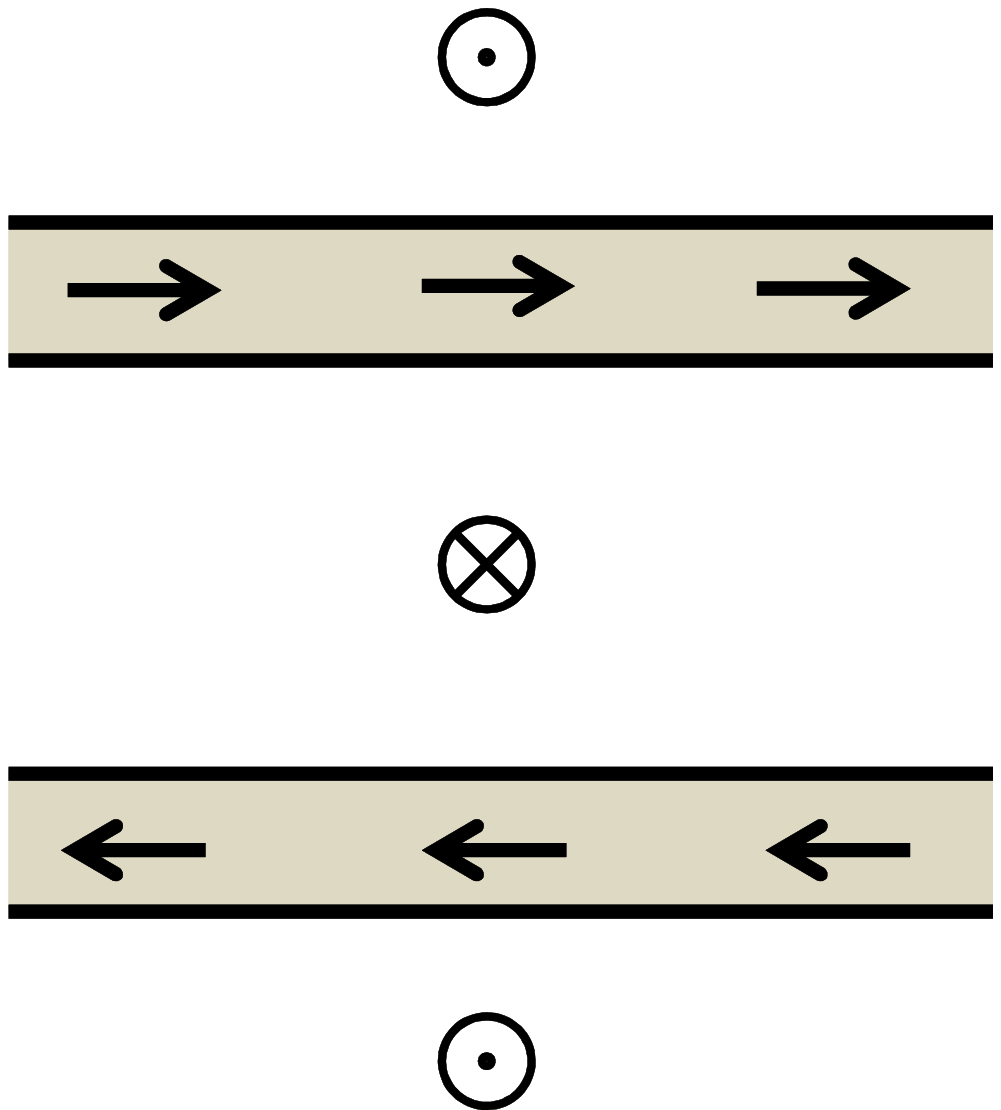}
\quad
\includegraphics[width=0.2\linewidth,keepaspectratio]{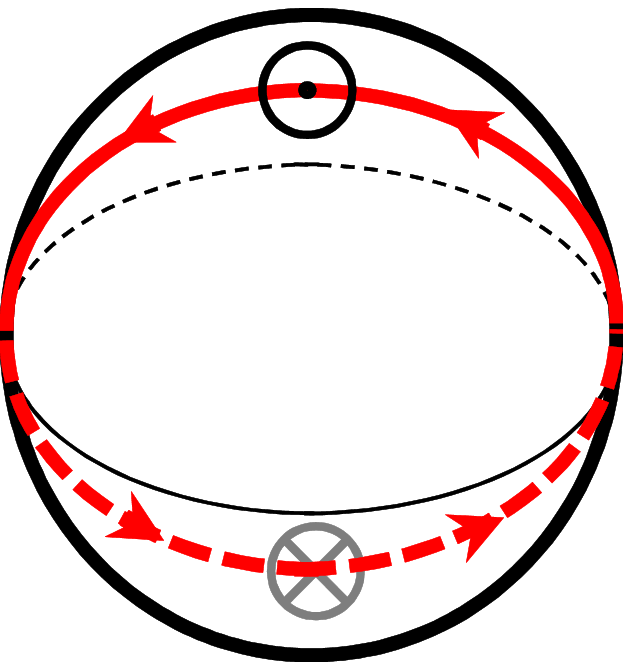}\\
(a)\hspace{3.5cm} (b)\hspace{3.5cm}
(c)\hspace{3.5cm} (d)

\caption{(a,b) Single domain wall and (c,d) a pair of a domain wall and 
an anti-domain wall in the ${\bf C}P^1$ model. 
(a) 
The wall is perpendicular to the $x^1$-axis.
The arrows denote points in the ${\bf C}P^1$.
(b) The ${\bf C}P^1$ target space. 
The north and south poles are denoted by 
$\odot$ and $\otimes$, respectively.
The path connecting them represents the map from 
the path in (a) along the $x^1$-axis 
in real space from $x^1 \to - \infty$ to $x^1 \to + \infty$. 
The path in the ${\bf C}P^1$ target space 
passes through one point on the equator, 
which is represented by $\leftarrow$ in (a) in this example. 
In general, the $U(1)$ zero mode is localized 
on the wall as a phase modulus.
(c) 
The phases of the wall and anti-wall
are  opposite to each other.
(d) The path represents the map from the wall and anti-wall 
configuration. 
}
\label{fig:brane-anti-brane-2d} 
\end{center}
\end{figure}

\section{Baby Skyrmions from wall-anti-wall annihilation}

Let us first construct a domain wall solution 
\cite{Kudryavtsev:1997nw,Harland:2007pb}.
For the existence of a domain wall, 
the FS term is not essential. 
We simply consider a small coupling $\kappa$. 
The model reduces to 
the massive ${\bf C}P^1$ model, 
which can be made supersymmetric by adding fermions 
\cite{Abraham:1992vb}. 
However, supersymmetry is not essential in our study. 

Then, a domain wall interpolating the two vacua 
$u=0$ and $u=\infty$ can be obtained as \cite{Abraham:1992vb}
\beq
 u_{\rm w} = e^{\mp m (x^1-x^1_0) + i \ph}  \label{eq:wall-sol}
\eeq
with the width $1/m$, 
where $\mp$ represents a wall and an anti-wall.
Here, $x^1_0$ and $\ph$ are real constants 
representing the position and phase 
of the (anti-)domain wall.
 The domain wall is mapped to 
a large circle starting from the north pole, 
denoted by $\odot$,
and ending at the south pole, 
denoted by $\otimes$,
in the ${\bf C}P^1$ target space.

\begin{figure}
\begin{center}
\includegraphics[width=0.2\linewidth,keepaspectratio]{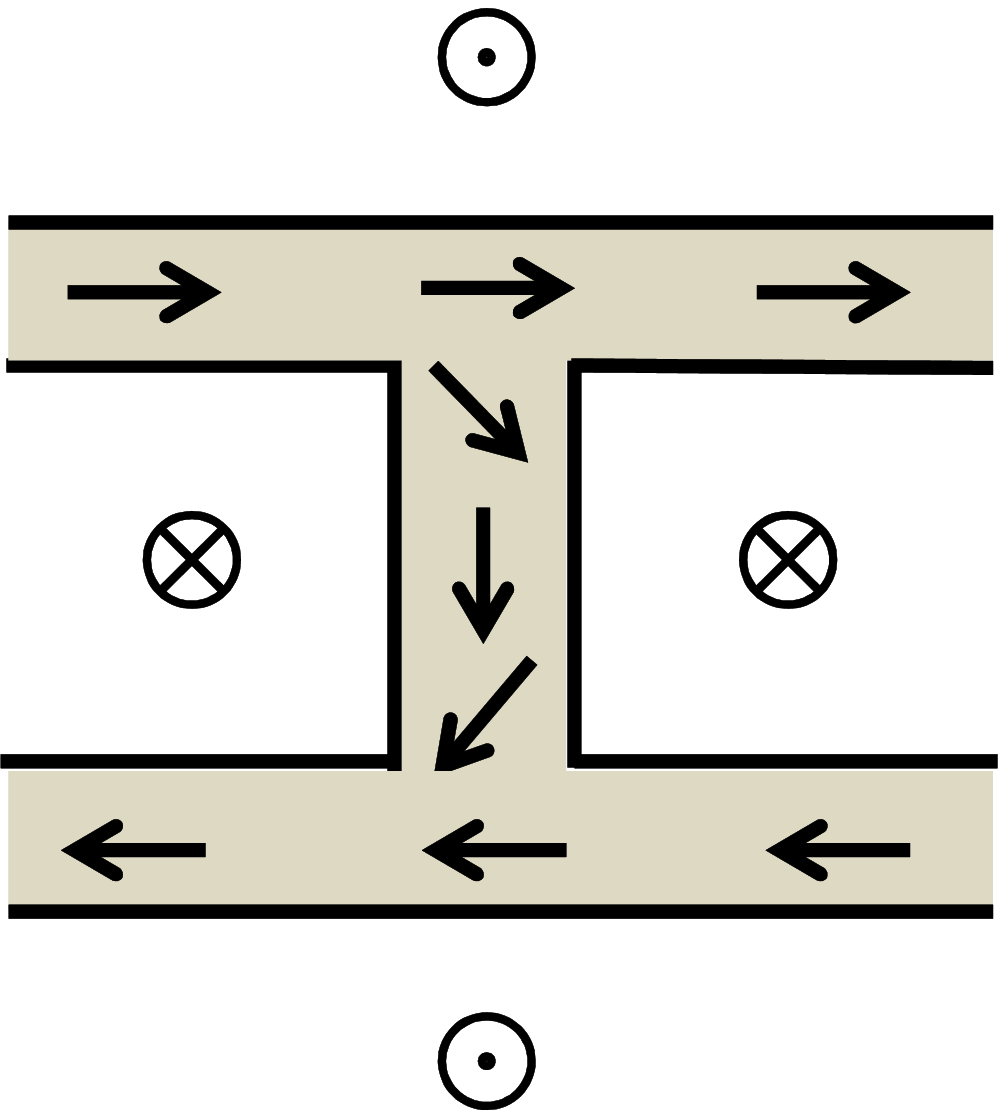}
\quad
\includegraphics[width=0.2\linewidth,keepaspectratio]{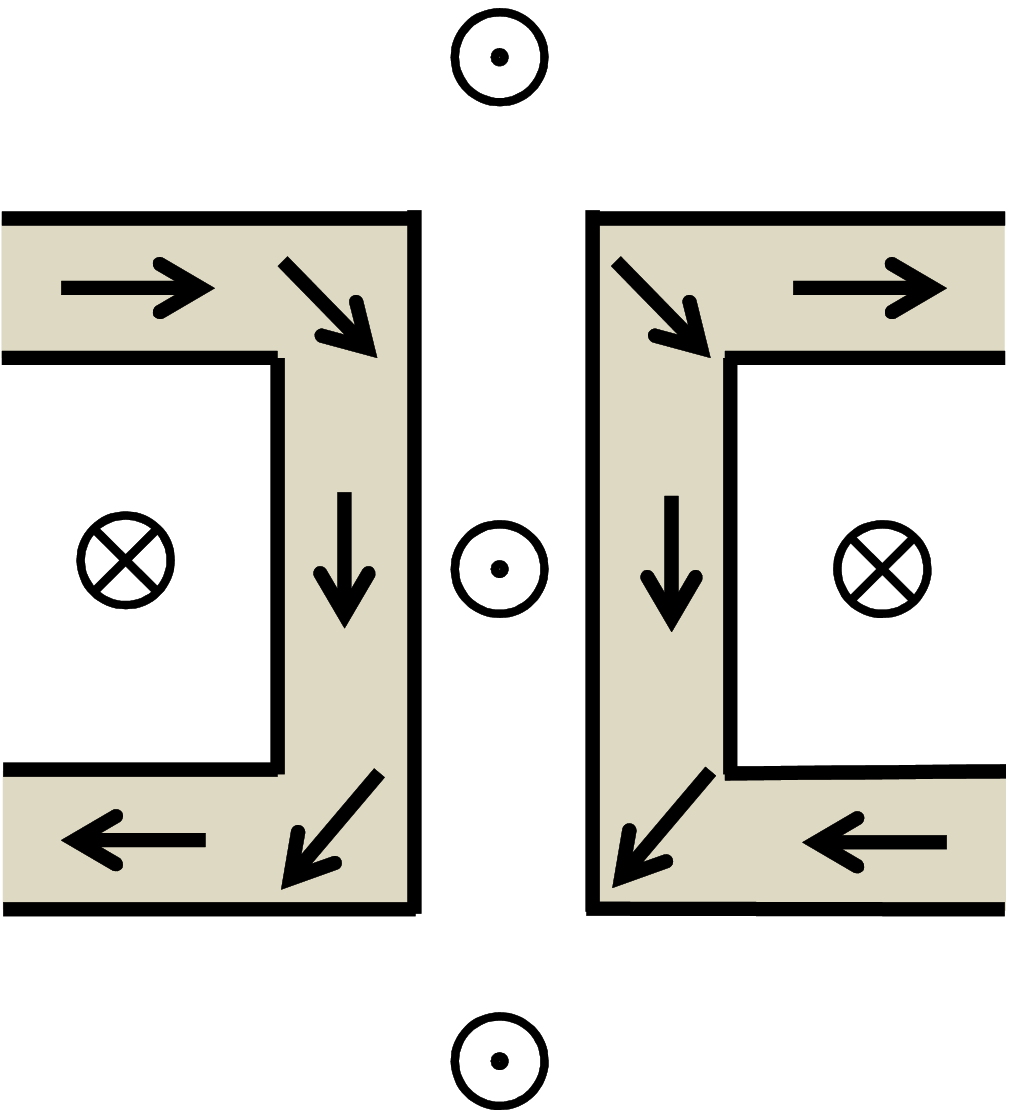}
\quad
\includegraphics[width=0.2\linewidth,keepaspectratio]{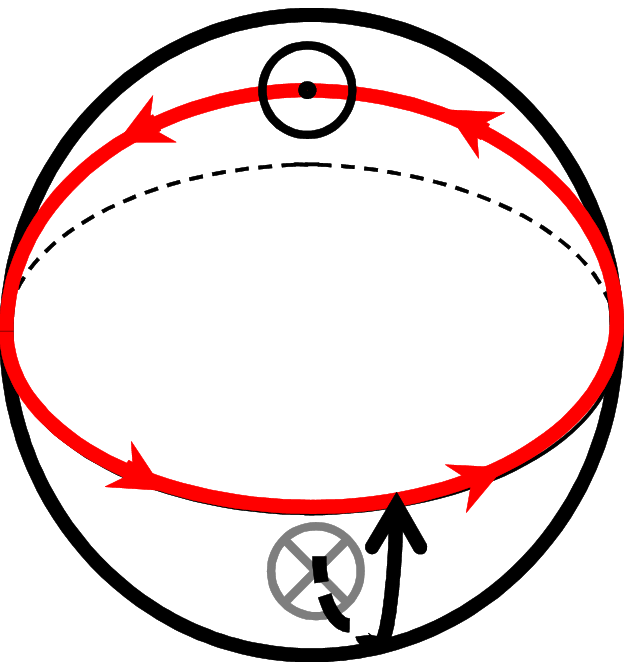}
\\
(a)\hspace{3cm} (b)\hspace{3cm} (c)\\

\medskip
\includegraphics[width=0.2\linewidth,keepaspectratio]{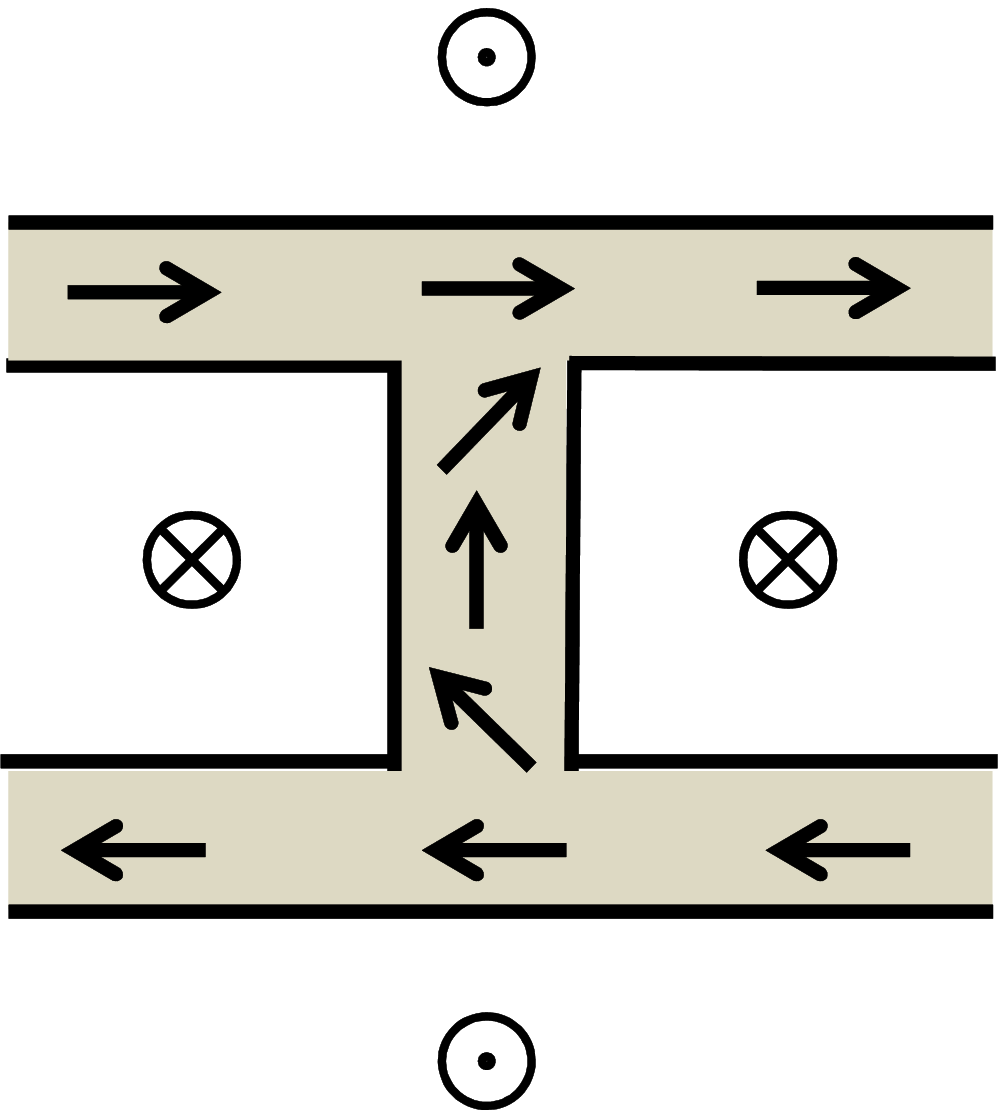}
\quad
\includegraphics[width=0.2\linewidth,keepaspectratio]{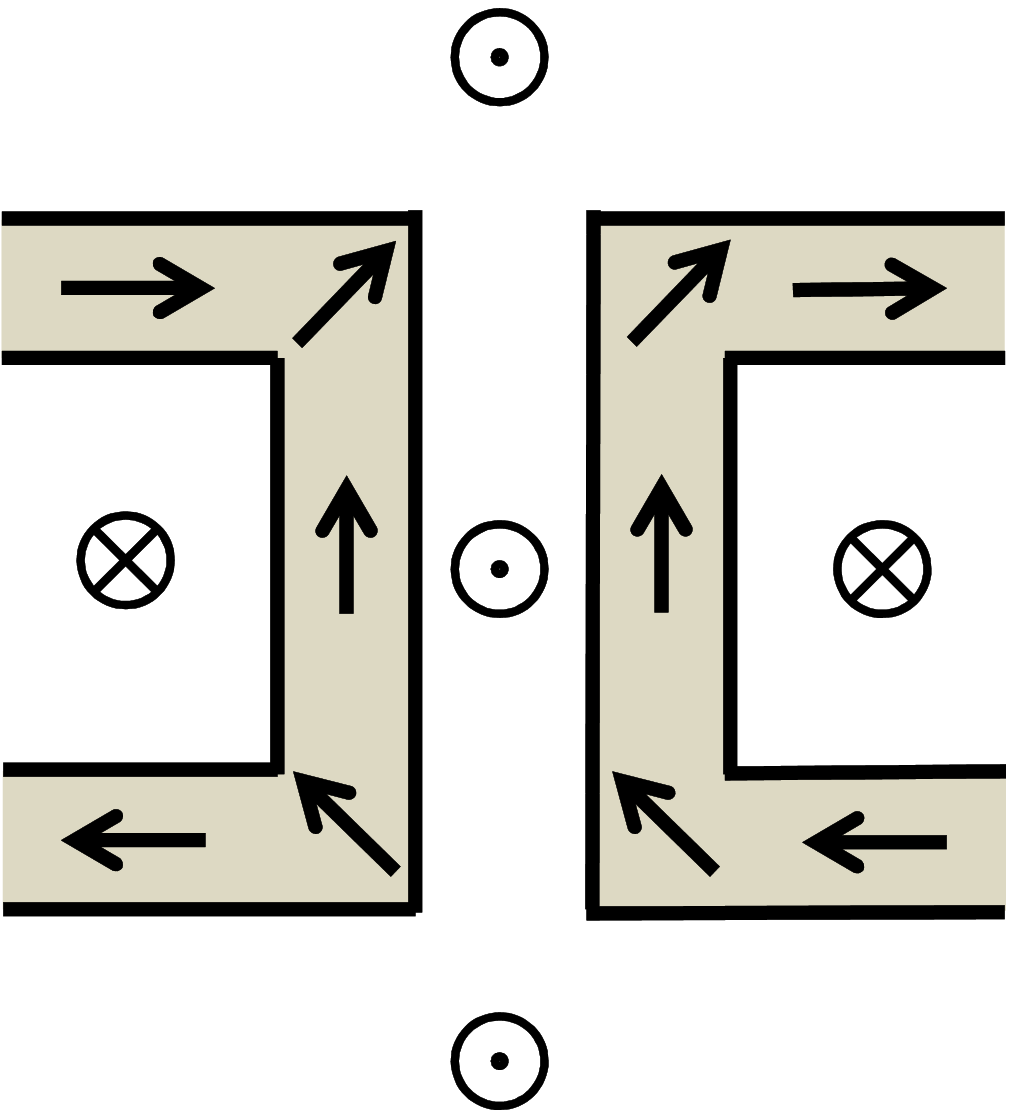}
\quad
\includegraphics[width=0.2\linewidth,keepaspectratio]{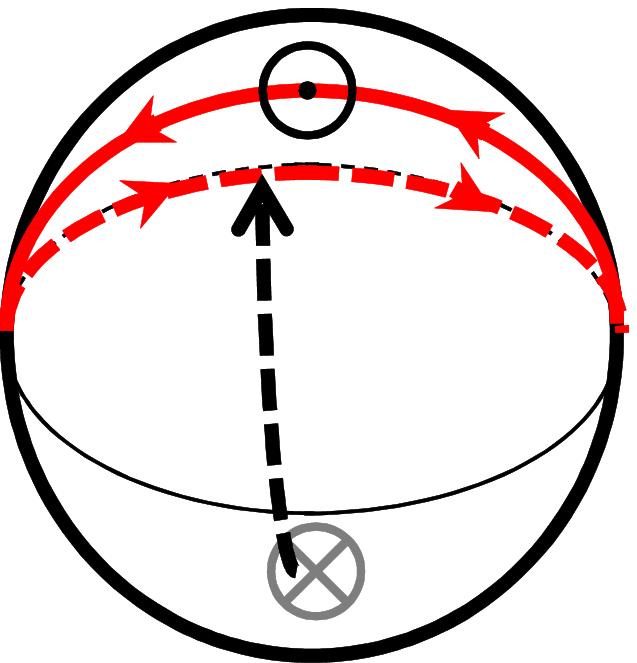}
\\
(d)\hspace{3cm} (e)\hspace{3cm} (f)
\caption{
Decaying processes of the wall and anti-wall.
(a,d) A bridge is created between the wall and the anti-wall. 
In this process, there are two possibilities of 
the ${\bf C}P^1$ structure 
along the bridge. 
(b,e) The upper and lower regions are connected 
by breaking the bridge. 
(c,f) Accordingly, the loop in the ${\bf C}P^1$ target space 
is unwound in two ways.
\label{fig:wall-anti-wall-annihilation} 
} 
\end{center}
\end{figure}

We consider a pair of a wall 
at $x^1=x^1_1$ 
and an anti-wall at $x^1 = x^1_2$. 
An appoximate solution valid at large distance, 
$x^1_2-x^1_1 \gg m^{-1}$, is obtained as
\beq
 u_{{\rm w}-{\rm aw}} 
= e^{- m (x^1-x^1_1) + i \ph_1} +  e^{+ m (x^1-x^1_2) + i \ph_2}.
\label{eq:wall-anti-wall-sol}
\eeq
Here, the phases $\ph_1$ and $\ph_2$ of the wall and 
anti-wall are considered to be opposite to each other,  
$\ph_1=\ph_2 +\pi$, 
as in Fig.~\ref{fig:brane-anti-brane-2d}(c).  
The configuration is mapped to a loop 
in the ${\bf C}P^1$ target space, 
as in Fig.~\ref{fig:brane-anti-brane-2d}(d).  
This configuration is unstable as  
it should end up with the vacuum state 
$\odot$.
In the decaying process, the loop is unwound from 
the south pole in the target space. 
The unwinding of the loop can be achieved in 
two topologically inequivalent processes, 
schematically shown in  
Fig.~\ref{fig:wall-anti-wall-annihilation}(c) and (f). 

In real space, 
at first, a bridge connecting the wall and anti-wall is created, 
as in Fig.~\ref{fig:wall-anti-wall-annihilation}(a) and (d).
Here, there exist two possibilities of the spin structure along the bridge, 
corresponding to the two inequivalent ways of 
the unwinding processes:
along the bridge in the $x^1$-direction, 
the spin rotates (a) anti-clockwise or (b) clockwise on the equator 
of the ${\bf C}P^1$ target space.
Let us label these two kinds of bridges 
by $\uparrow$ and $\downarrow$. 
Second, the bridge is broken into two pieces, 
as in Fig.~\ref{fig:wall-anti-wall-annihilation}(b) and (e)
between which the vacuum state $\odot$ is filled between them. 
Let us again label these two kinds of holes 
by $\uparrow$ and $\downarrow$.
In either case, the two regions separated by the domain walls 
are connected through a hole created by the decay of the walls.
Once created, these holes grow with reducing the wall energy.
\begin{figure}
\begin{center}
\includegraphics[width=0.2\linewidth,keepaspectratio]{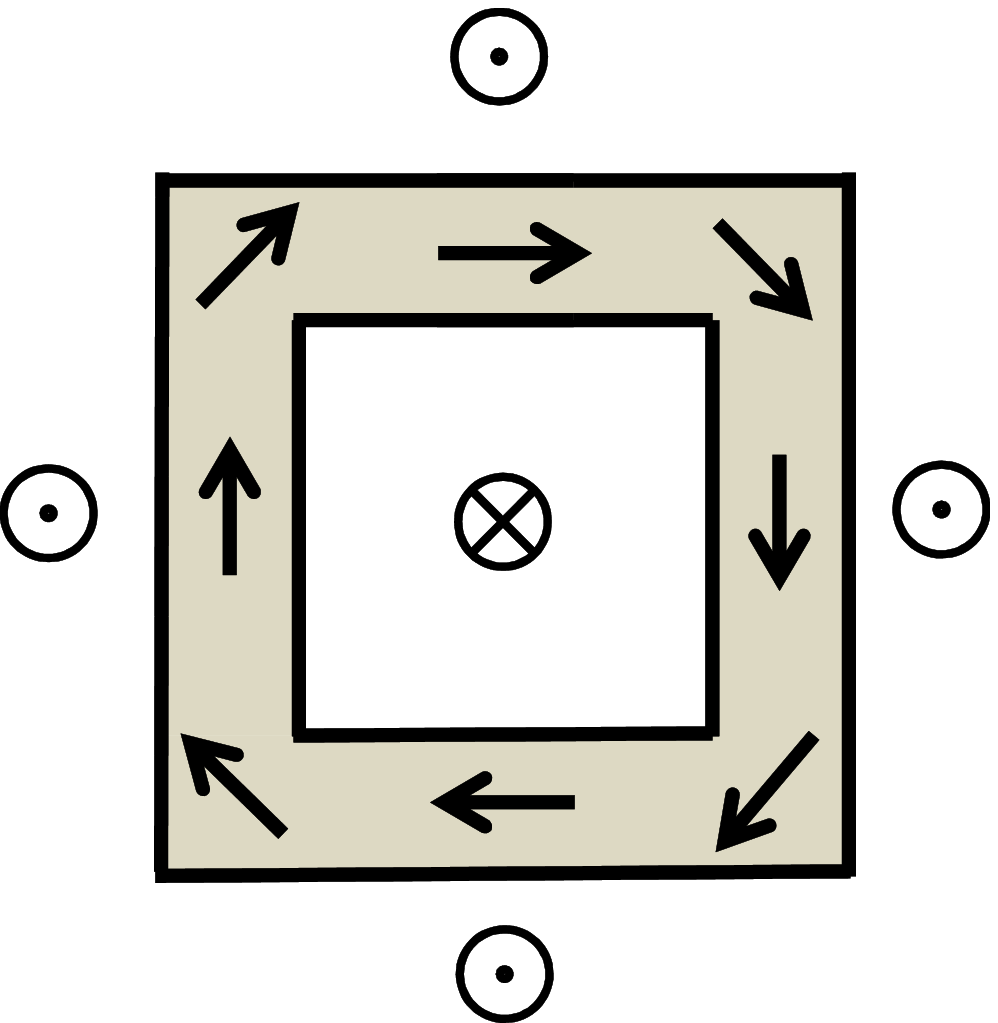}\quad
\includegraphics[width=0.2\linewidth,keepaspectratio]{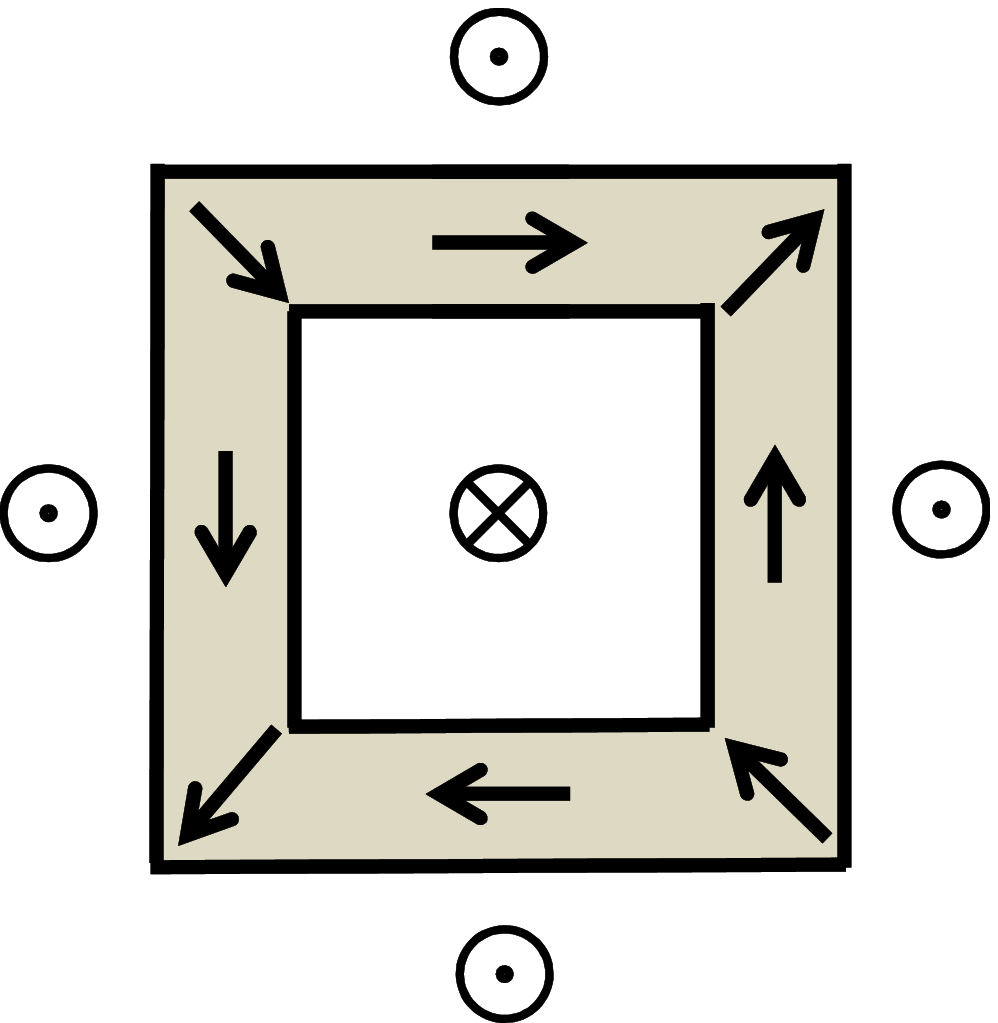}\quad
\includegraphics[width=0.2\linewidth,keepaspectratio]{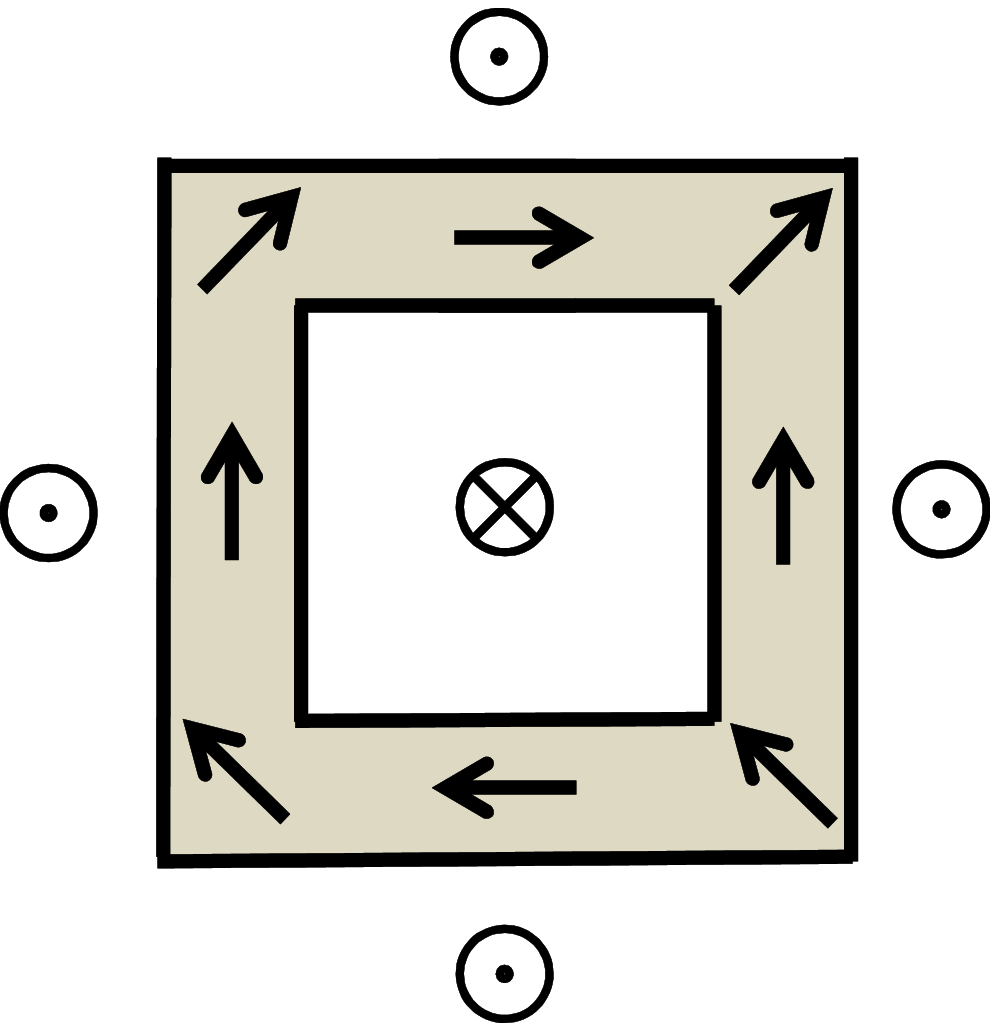}\quad
\includegraphics[width=0.2\linewidth,keepaspectratio]{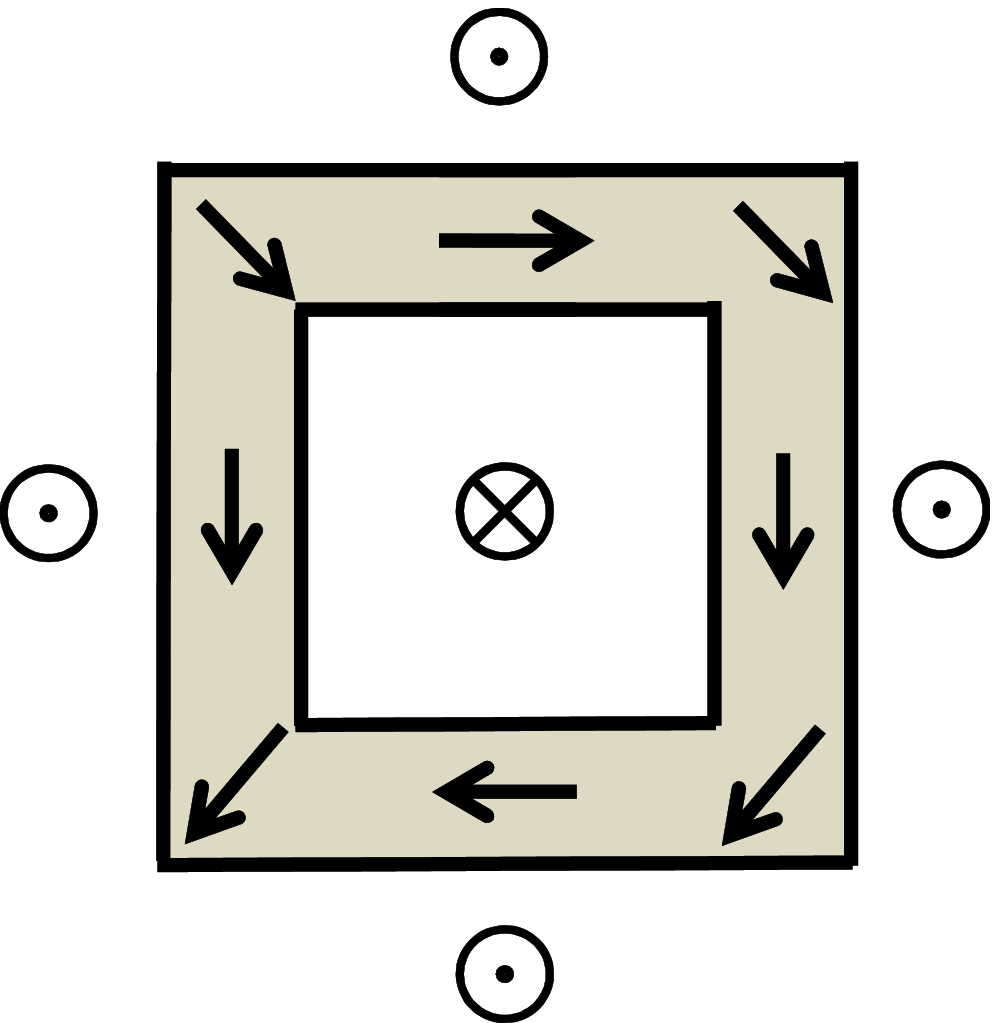}

\hspace{2cm}
(a)\hspace{3cm} (b)\hspace{3cm}
(c)\hspace{3cm} (d)\hspace{3cm} 

\end{center}
\caption{Stable and unstable rings. 
(a,b) Stable rings. The phase winds once 
(the winding number is $\pm 1$) along the rings. 
Total configurations are baby skyrmions with 
a non-trivial element, $\pm 1$, of the second homotopy group, $\pi_2$.
(c,d) Unstable rings. 
The phase does not wind  
(the winding number is $0$) along the rings. 
They decay into ground state (up pseudo-spin).
} 
\label{fig:winding} 
\end{figure}

Several holes are created during the entire decaying process.
Let us focus a pair of two nearest holes. 
One can find a ring of a domain wall between the holes, 
as shown in Fig.~\ref{fig:winding}.
Here, since there exist two kinds of holes ($\uparrow$ and $\downarrow$), 
there exist four possibilities of the rings,  
(a) $\uparrow\downarrow$, (b) $\downarrow\uparrow$, 
(c) $\uparrow\uparrow$, and (d) $\downarrow\downarrow$  
in Fig.~\ref{fig:winding}. 
Clearly, the rings of types (c) of (d) can decay and 
end up with the vacuum state $\odot$. 
However, the decay of the rings of types 
(a) and (b) is topologically forbidden,  
because of a nontrivial winding of the spin along the rings. 
What are these topologically protected rings of a domain wall?
They are nothing but baby skyrmions.
The solutions can be written as ($\omega \equiv x^1+ix^2$)
$u = u_0 = \lambda/(\omega-\omega_0)$ and  $u = \bar u_0$ 
for a baby skyrmion and an anti-baby skyrmion, respectively, 
where $\omega_0 \in {\bf C}$ represents 
the positions of the lump. 
Here, $\lambda \in {\bf C}^*$
where the size $|\lambda|$ is fixed by the balance between 
the potential term and the FS term, 
while $U(1)$ orientation $\arg \lambda$ 
is a modulus of the baby skyrmion. 
In fact,
these configurations can be shown to have a nontrivial winding 
in the second homotopy group $\pi_2 ({\bf C}P^1)\simeq {\bf Z}$; 
(a) and (b) belong to, respectively, 
$+1$ and $-1$ of $\pi_2 ({\bf C}P^1)$.

\begin{figure}[h]
\begin{center}
\includegraphics[width=0.4\linewidth,keepaspectratio]{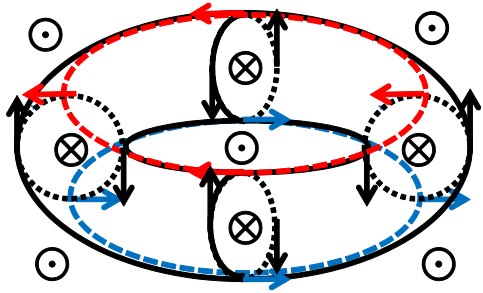}
\end{center}
\caption{An untwisted baby srkyme-ring.\label{fig:vortex-ring}
} 
\end{figure}

In $d=3+1$ dimensions, 
domain walls have two spatial dimensions 
in their world-volume. 
When a domain wall-pair decay occurs, 
there appear two-dimensional 
holes, which are labeled as 
$\downarrow$ or $\uparrow$ in 
Fig.~\ref{fig:wall-anti-wall-annihilation} (b) and (e), 
respectively.
Along the boundary of these two kinds of holes, 
there appear baby skyrme strings, which generally create 
closed baby skyrme strings, {\it i.e.}, baby skyrme rings,  
as in Fig.~\ref{fig:vortex-ring}.  
This process can be numerically verified \cite{Takeuchi:2011}.
These rings are unstable to decay 
into the fundamental excitations in the end.

\section{Creating Knots}

\begin{figure}
\begin{center}
\includegraphics[width=0.4\linewidth,keepaspectratio]{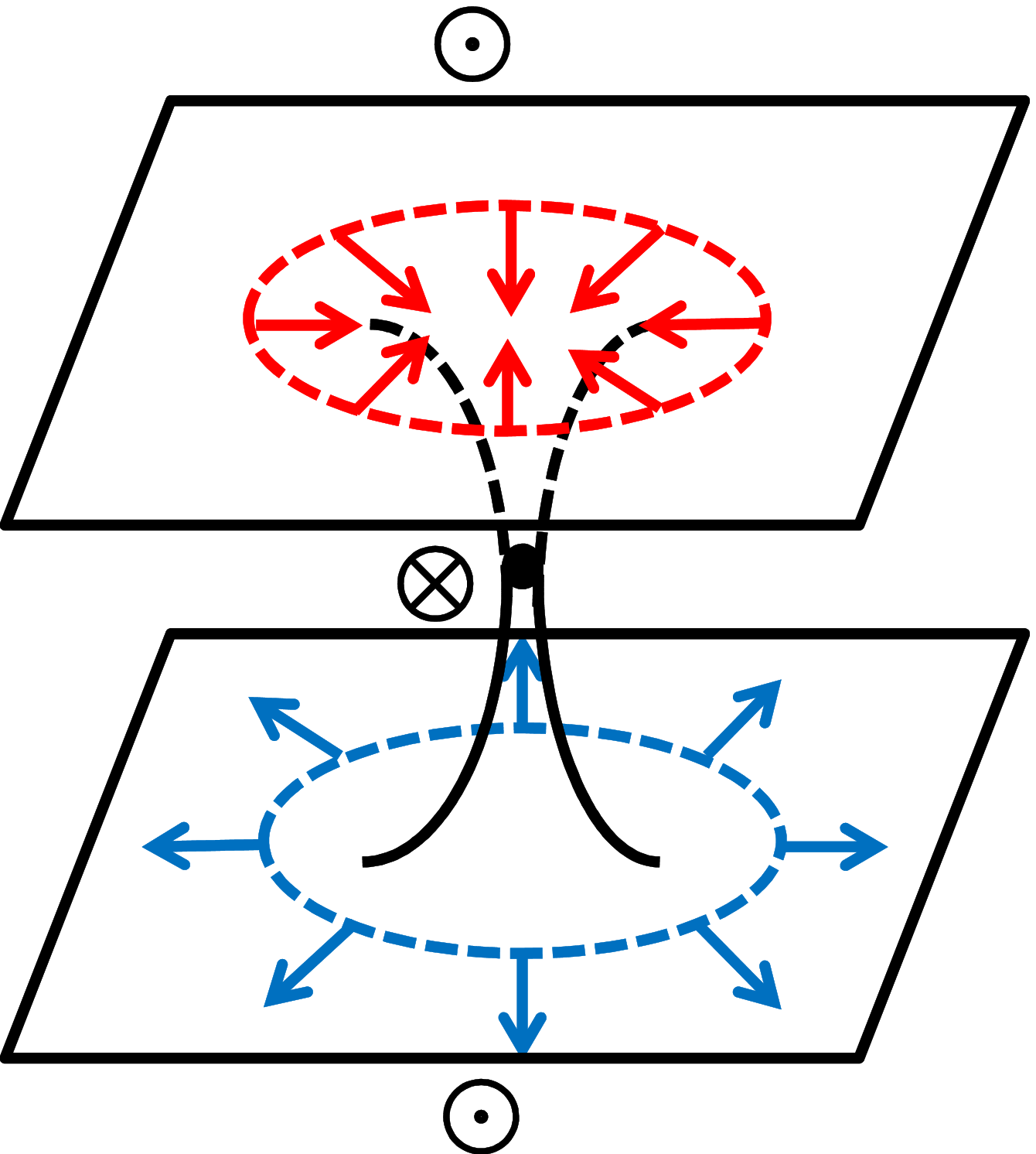}
\quad
\includegraphics[width=0.35\linewidth,keepaspectratio]{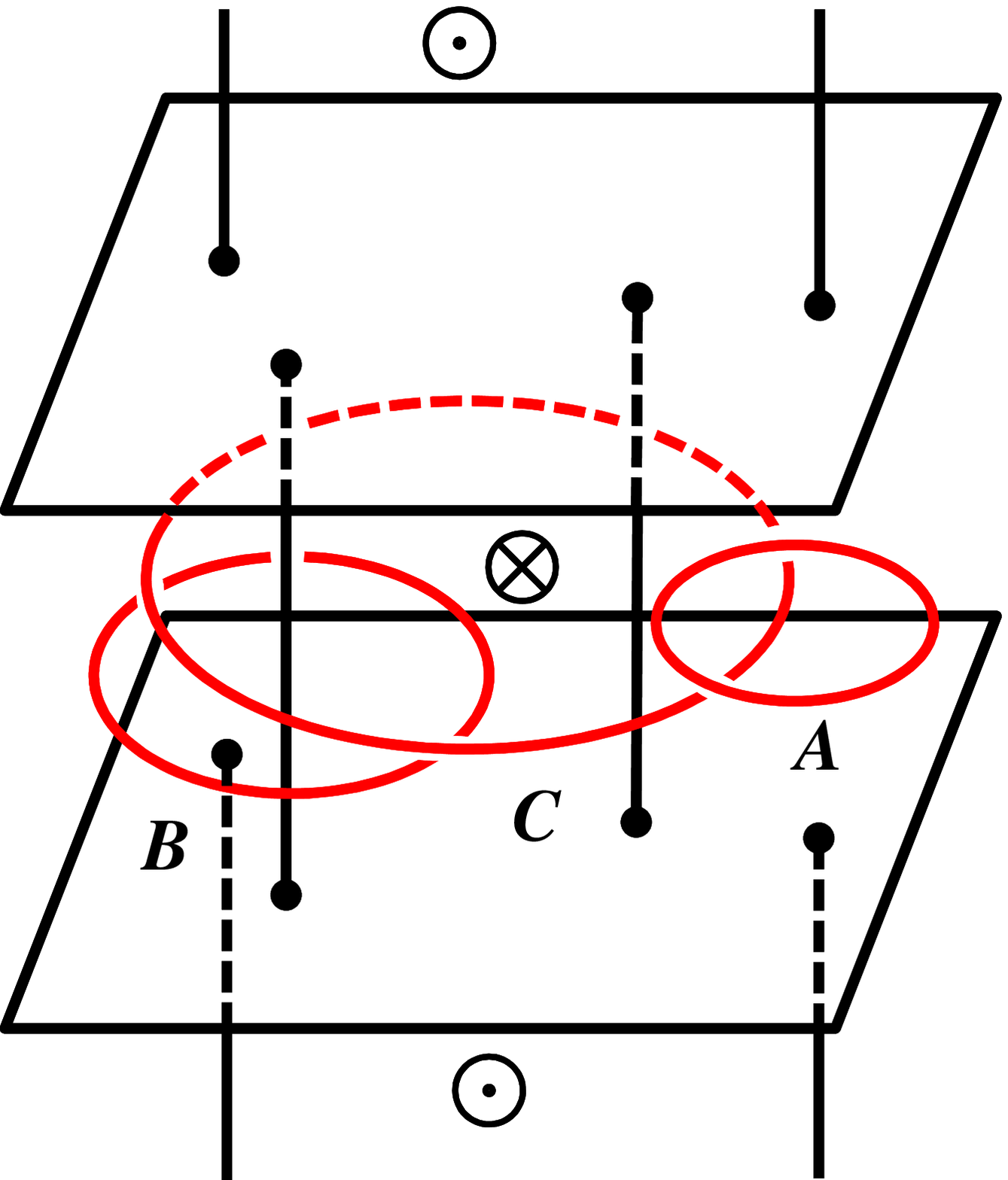}\\
(a) \hs{50} (b)\\
\end{center}
\caption{(a) A pair of a domain wall and 
an anti-domain wall stretched by a 
string (vortex).
The domain walls are perpendicular to the $x^3$-axis, 
and a baby skyrme string along the $x^3$-axis
is stretched between the domain walls.
The arrows denote points in the ${\bf C}P^1$ target space.  
(b)
Loops in the wall-vortex systems. 
While the loop A yields an untwisted baby skyrme ring 
in Fig.~\ref{fig:vortex-ring}, 
the loop B (C) yields 
a baby skyrme ring twisted once (twice)
with the Hopf charge one (two).
\label{fig:brane-anti-brane-with-string} 
} 
\end{figure}

Here, we discuss the approach to stabilizing baby skyrme rings.
To this end, 
we place a baby skyrme string 
stretching between the domain wall and the anti-domain wall, 
as in Fig.~\ref{fig:brane-anti-brane-with-string}. 

\begin{figure}[t]
\begin{center}
\includegraphics[width=0.4\linewidth,keepaspectratio]{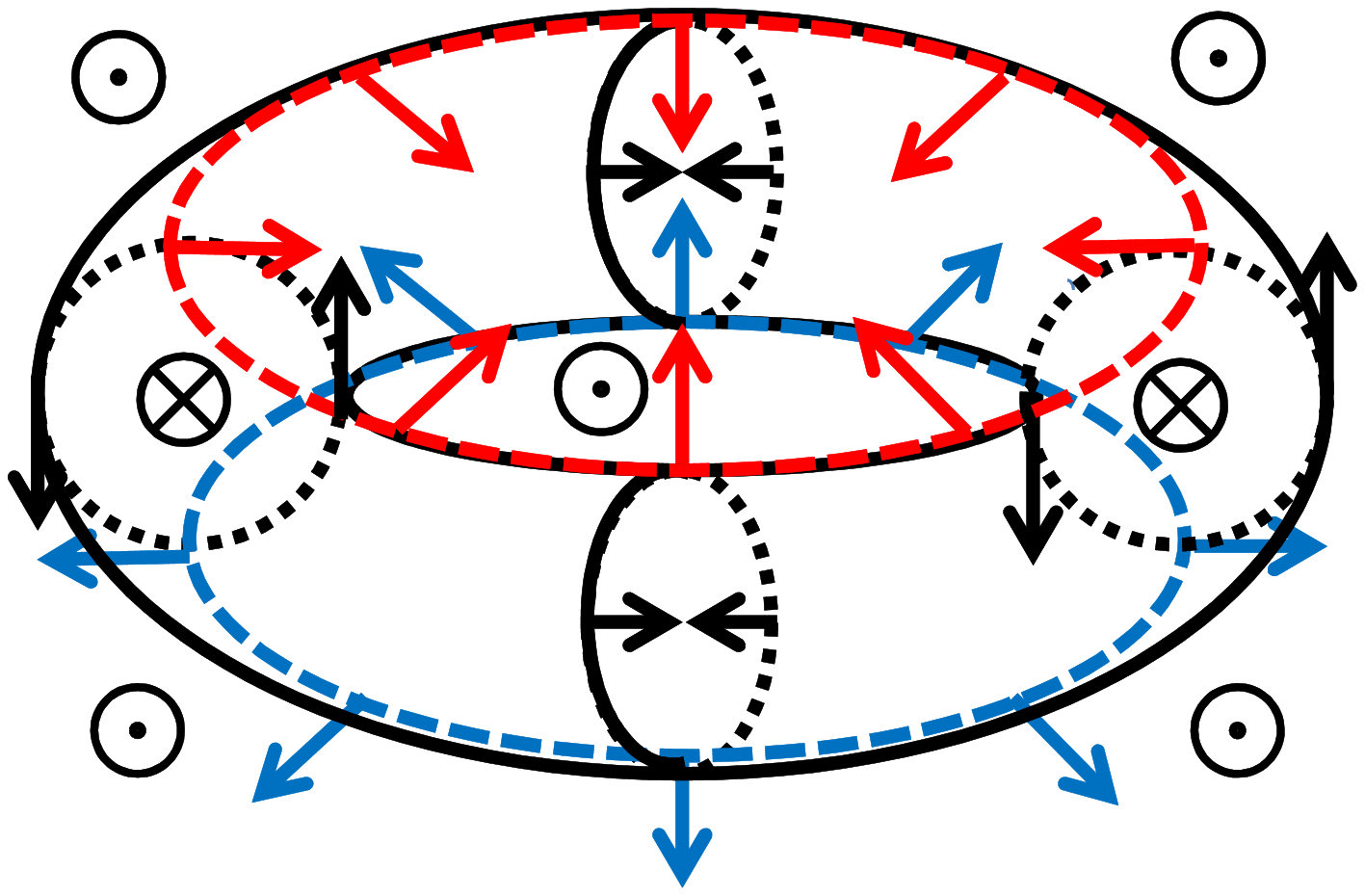}
\quad
\includegraphics[width=0.4\linewidth,keepaspectratio]{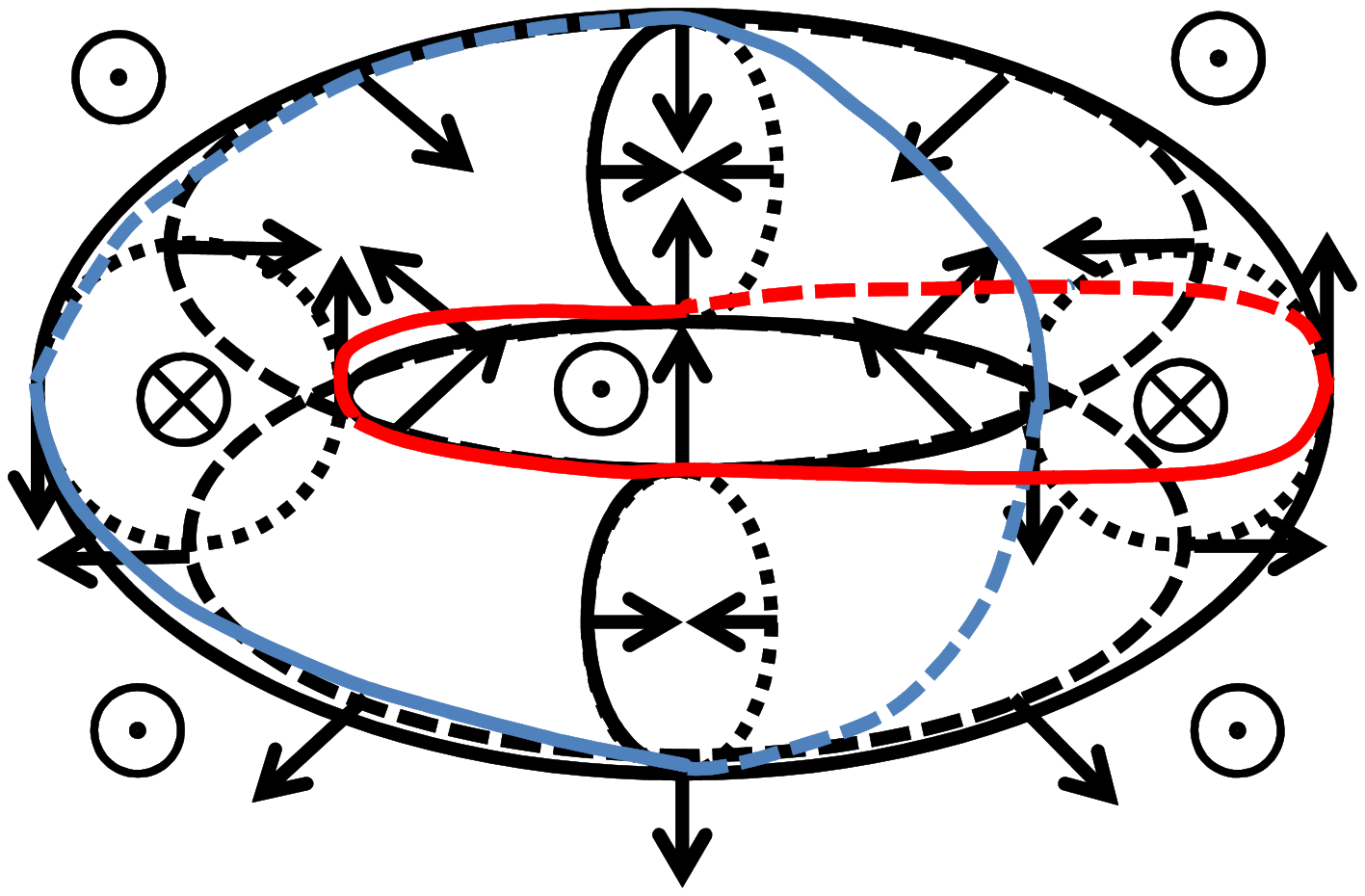}
\\
(a)
\hs{60}
(b)
\end{center}
\caption{A knot created after a wall--anti-wall pair annihilation.
(a) 
The torus divides the regions of 
$\odot$ and $\otimes$.  
The vertical section of the torus in the $x^1$-$x^2$ plane  
is a pair of a baby skyrmion and an anti-baby skyrmion. 
While they rotate along the $x^3$-axis,  
their phases are twisted and connected to each other 
at the $\pi$ rotation. 
This configuration is different from that 
of a closed and untwisted baby skyrme srting in Fig.~\ref{fig:vortex-ring}.
(b) The preimages of $\uparrow$ and $\downarrow$ make a link.
\label{fig:knot}
} 
\end{figure}

First, we consider $k$ baby skyrme strings 
ending on a domain wall. 
Such a configuration was constructed in 
\cite{Gauntlett:2000de,Isozumi:2004vg}
in the absence of the FS term ($z\equiv x^2+ix^3$)
\beq
u_{\rm w-v} 
= e^{\pm m (x^1 - x^1_0) + i \ph_0}Z(z), 
 \quad 
 Z(z) = \sum_{i=1}^k {\lambda_i \over z- z_i}.
\label{eq:D-brane-soliton}
\eeq
This solution precisely coincides with 
a BIon \cite{Callan:1997kz} 
in the Dirac-Born-Infeld action on a D2-brane, 
and so, this was referred to as a ``D-brane soliton" 
\cite{Gauntlett:2000de}. 
The wall surface in the above solution is logarithmically bent. 
We can also place the strings on the both sides of the wall 
\cite{Isozumi:2004vg}
\begin{equation}
Z(z)
 = {\prod_{j=1}^{k_+} (z - z_j^+)
\slash
\prod_{i=1}^{k_-} (z - z_i^-)}, 
\label{eq:D-brane-soliton2}
\end{equation}
where $z_i^\pm$ and $k_\pm$  
denote the positions and numbers of 
strings extending to $x^1 \to \pm \infty$, 
respectively. 
If the numbers of strings coincide 
on both sides, {\it i.e.}, $k_1=k_2$ , 
the wall surface is asymptotically flat.

With regard to $\kappa$, the above solution should be modified; 
however, we consider a regime with a small $\kappa$ 
so that the modifications to the above solution 
can be neglected.

Now let us consider the strings stretched between domain walls 
in Fig.~\ref{fig:brane-anti-brane-with-string}. 
An approximate solution for 
a wall with the phase $\ph_1$  at $x^1=x^1_1$ 
and an anti-wall with the phase $\ph_2$ at $x^1=x^1_2$ 
is obtained from Eq.~(\ref{eq:wall-anti-wall-sol}) as 
\beq
u_{\rm w-v-aw} 
= (e^{- m (x^1 - x^1_1) + i \ph_1} 
 + e^{+ m (x^1 - x^1_2) + i \ph_2}) 
Z(z), 
\eeq
with $Z(z)$ in Eq.~(\ref{eq:D-brane-soliton}) 
for $k$ stretched strings, 
or Eq.~(\ref{eq:D-brane-soliton2}) 
for $k_2$ stretched strings 
and $k_1$ strings attached from outside.
In our case, we take $\ph_1 = \ph_2 +\pi$.

As in the case without a stretched string,
the configuration itself is unstable to decay, 
and baby skyrme rings are created. 
A ring is not twisted if it does not enclose
the stretched strings, 
as the loop A in Fig.~\ref{fig:brane-anti-brane-with-string}(b).
However, if a ring encloses $n$ stretched strings,  
as the loops B and C in Fig.~\ref{fig:brane-anti-brane-with-string}(b), 
it is twisted $n$ times.
A baby skyrme ring twisted once is shown in Fig.~\ref{fig:knot}(a). 
The vertical section of the torus in the $x^1$-$x^2$-plane  
is a pair of a baby skyrmion and an anti-baby skyrmion. 
Moreover, the presence of the stretched string 
implies that the phase winds anti-clockwise along the loops, 
as is indicated by the arrows on the top and bottom of the torus 
in Fig.~\ref{fig:knot}(a). 
When the baby skyrmions in the pair rotate along the $x^1$-axis, 
their phases are twisted and connected to each other 
at the $\pi$ rotation. 
This configuration is nothing but a Hopfion, 
which is stable as discussed in \cite{Gladikowski:1996mb}.
One can calculate the Hopf charge,  
but there is a simpler way to confirm that the configuration is a 
Hopfion.
A preimage of a point on ${\bf C}P^1$ is a loop 
in real space.
When two loops of the preimages of two arbitrary points on ${\bf C}P^1$ 
have a linking number $n$, 
the configuration has a Hopf charge $n$.
In Fig.~\ref{fig:knot}(b), we plot the preimages of 
$\uparrow$ and $\downarrow$, 
which are linked with a linking number one. 
Thus, we obtain a knot soliton with a Hopf charge of one 
(a Hopfion). 
Similarly, a created ring enclosing $n$ stretched strings 
yields a knot soliton with a Hopf charge of $n$.

In summary, when a pair of a domain wall and an anti-domain wall 
annihilates in the FS model with the potential term, 
(anti-)baby skyrmions are created in $d=2+1$ dimensions. 
In $d=3+1$ dimensions, there appear baby skyrme-strings. 
When a string is stretched between the wall pair, 
the baby skyrme-ring that encircles it  
becomes a twisted baby skymre ring, which 
is a Hopfion. 
This mechanism is the first proposal for 
creating knot solitons.

{\bf Acknowledgements}

This work is supported in part by 
Grant-in Aid for Scientific Research (No.~23740198) 
and by the ``Topological Quantum Phenomena'' 
Grant-in Aid for Scientific Research 
on Innovative Areas (No.~23103515)  
from the Ministry of Education, Culture, Sports, Science and Technology 
(MEXT) of Japan.

\end{document}